\documentclass[runningheads]{llncs}
\usepackage{graphicx}
\usepackage{multirow}
\usepackage{booktabs}
\usepackage{bbding}
\usepackage{hyperref, color}

\usepackage{colortbl}
\usepackage[table,xcdraw]{xcolor}
\usepackage{pifont}
\usepackage{latexsym, amssymb}
\usepackage{amsmath}

\begin{document}
\title{
S$^2$ME: Spatial-Spectral Mutual Teaching and Ensemble Learning for Scribble-supervised Polyp Segmentation
}
\titlerunning{S$^2$ME: Spatial-Spectral Mutual Teaching and Ensemble Learning}
%
\author{An Wang\inst{1} 
\and Mengya Xu\inst{2} 
\and Yang Zhang\inst{1,3}
\and Mobarakol Islam\inst{4} 
\and Hongliang Ren\inst{1,2} 
\thanks{Corresponding author.}}

%
\authorrunning{A. Wang et al.}
%
\institute{Dept. of Electronic Engineering, Shun Hing Institute of Advanced Engineering (SHIAE), The Chinese University of Hong Kong, Hong Kong SAR, China
\and Dept. of Biomedical Engineering, National University of Singapore, Singapore
\and School of Mechanical Engineering, Hubei University of Technology, Wuhan, China
\and Dept. of Medical Physics and Biomedical Engineering, Wellcome/EPSRC Centre for Interventional and Surgical Sciences (WEISS), University College London, London, UK\\
\email{wa09@link.cuhk.edu.hk, mengya@u.nus.edu, yzhangcst@hbut.edu.cn, mobarakol.islam@ucl.ac.uk, hlren@ee.cuhk.edu.hk}
}
\maketitle              
\begin{abstract}
Fully-supervised polyp segmentation has accomplished significant triumphs over the years in advancing the early diagnosis of colorectal cancer. However, label-efficient solutions from weak supervision like scribbles are rarely explored yet primarily meaningful and demanding in medical practice due to the expensiveness and scarcity of densely-annotated polyp data. Besides, various deployment issues, including data shifts and corruption, put forward further requests for model generalization and robustness. 
To address these concerns, we design a framework of \textbf{S}patial-\textbf{S}pectral Dual-branch \textbf{M}utual Teaching and Entropy-guided Pseudo Label \textbf{E}nsemble Learning (S$^2$ME). 
Concretely, for the first time in weakly-supervised medical image segmentation, we promote the dual-branch co-teaching framework by leveraging the intrinsic complementarity of features extracted from the spatial and spectral domains and encouraging cross-space consistency through collaborative optimization. 
Furthermore, to produce reliable mixed pseudo labels, which enhance the effectiveness of ensemble learning, we introduce a novel adaptive pixel-wise fusion technique based on the entropy guidance from the spatial and spectral branches. Our strategy efficiently mitigates the deleterious effects of uncertainty and noise present in pseudo labels and surpasses previous alternatives in terms of efficacy.
Ultimately, we formulate a holistic optimization objective to learn from the hybrid supervision of scribbles and pseudo labels. 
Extensive experiments and evaluation on four public datasets demonstrate the superiority of our method regarding in-distribution accuracy, out-of-distribution generalization, and robustness, highlighting its promising clinical significance. 
Our code is available at \url{https://github.com/lofrienger/S2ME}.
\end{abstract}
\section{Introduction}
\label{sec:intro}

Colorectal cancer is a leading cause of cancer-related deaths worldwide~\cite{ali2023multi}. Early detection and efficient diagnosis of polyps, which are precursors to colorectal cancer, is crucial for effective treatment. Recently, deep learning has emerged as a powerful tool in medical image analysis, prompting extensive research into its potential for polyp segmentation. 
The effectiveness of deep learning models in medical applications is usually based on large, well-annotated datasets, which in turn necessitates a time-consuming and expertise-driven annotation process. 
This has prompted the emergence of approaches for annotation-efficient weakly-supervised learning in the medical domain with limited annotations like points~\cite{he2022intra}, bounding boxes~\cite{li2022domain}, and scribbles~\cite{luo2022scribble}. Compared with other sparse labeling methods, scribbles allow the annotator to annotate arbitrary shapes, making them more flexible than points or boxes~\cite{lin2016scribblesup}. Besides, scribbles provide a more robust supervision signal, which can be prone to noise and outliers~\cite{cinbis2016weakly}. Hence, this work investigates the feasibility of conducting polyp segmentation using scribble annotation as supervision.
The effectiveness of medical applications during in-site deployment depends on their ability to generalize to unseen data and remain robust against data corruption. Improving these factors is crucial to enhance the accuracy and reliability of medical diagnoses in real-world scenarios~\cite{wang2022rethinking,xu2021class,xu2021learning}. Therefore, we comprehensively evaluate our approach on multiple datasets from various medical sites to showcase its viability and effectiveness across different contexts.

Dual-branch learning has been widely adopted in annotation-efficient learning to encourage mutual consistency through co-teaching. 
While existing approaches are typically designed for learning in the spatial domain~\cite{wu2021semi,valvano2021learning,zhang2022cyclemix,zhang2022shapepu}, a novel spatial-spectral dual-branch structure is introduced to efficiently leverage domain-specific complementary knowledge with synergistic mutual teaching. 
Furthermore, the outputs from the spatial-spectral branches are aggregated to produce mixed pseudo labels as supplementary supervision. 
Different from previous methods, which generally adopt the handcrafted fusion strategies~\cite{luo2022scribble}, we design to aggregate the outputs from spatial-spectral dual branches with an entropy-guided adaptive mixing ratio for each pixel. 
Consequently, our incorporated tactic of pseudo-label fusion aptly assesses the pixel-level ambiguity emerging from both spatial and frequency domains based on their entropy maps, thereby allocating substantially assured categorical labels to individual pixels and facilitating effective pseudo label ensemble learning.

\begin{figure}[!h]
\centering
\includegraphics[width=0.9\textwidth]{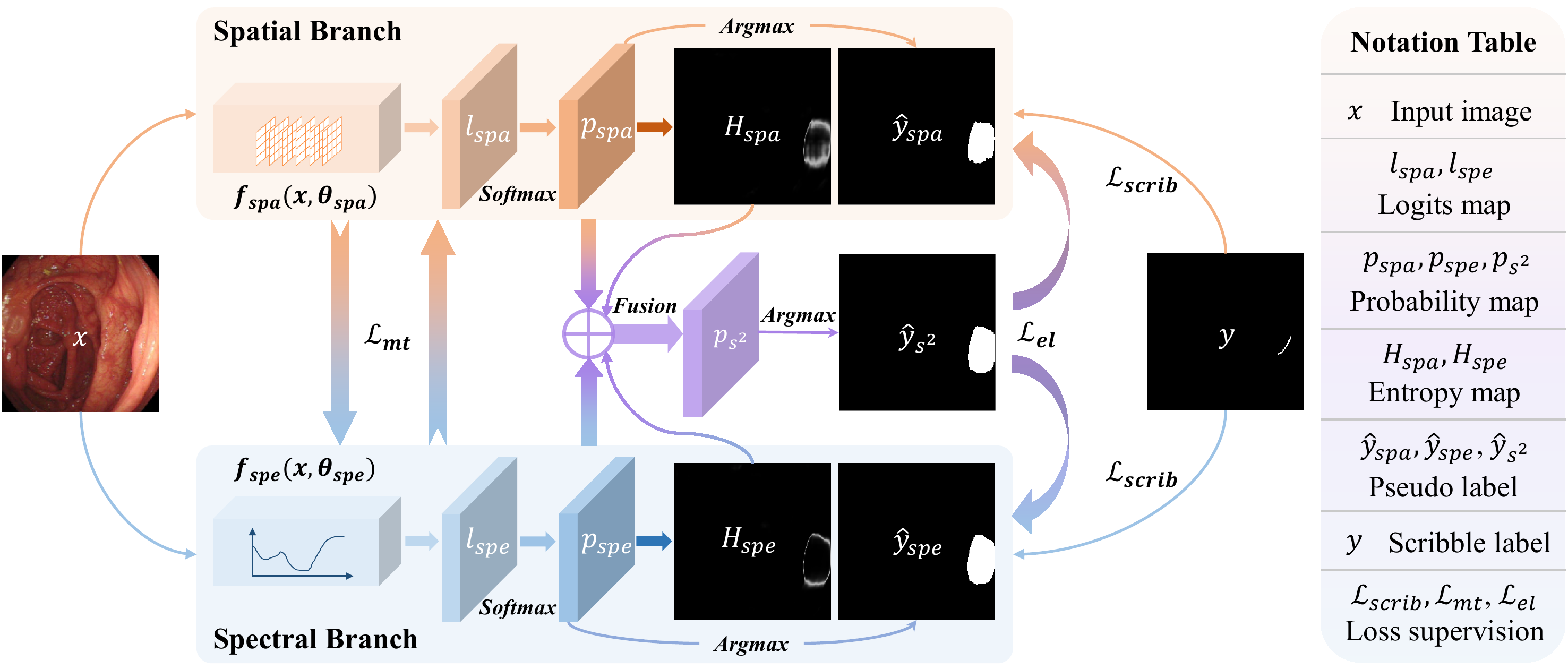}
\caption{Overview of our \textbf{S}patial-\textbf{S}pectral Dual-branch \textbf{M}utual Teaching and Pixel-level Entropy-guided Pseudo Label \textbf{E}nsemble Learning (S$^2$ME) for scribble-supervised polyp segmentation. 
Spatial-spectral cross-domain consistency is encouraged through mutual teaching. 
High-quality mixed pseudo labels are generated with pixel-level guidance from the dual-space entropy maps, ensuring more reliable supervision for ensemble learning.
}
\label{fig:overall}
\centering
\end{figure}

\subsubsection{Contributions.}
Overall, the contributions of this work are threefold: 
First, we devise a spatial-spectral dual-branch structure to leverage cross-space knowledge and foster collaborative mutual teaching. To our best knowledge, this is the first attempt to explore the complementary relations of the spatial-spectral dual branch in boosting weakly-supervised medical image analysis. Second, we introduce the pixel-level entropy-guided fusion strategy to generate mixed pseudo labels with reduced noise and increased confidence, thus enhancing ensemble learning. Lastly, our proposed hybrid loss optimization, comprising scribbles-supervised loss, mutual training loss with domain-specific pseudo labels, and ensemble learning loss with fused-domain pseudo labels, facilitates obtaining a generalizable and robust model for polyp image segmentation. An extensive assessment of our approach through the examination of four publicly accessible datasets establishes its superiority and clinical significance.

    
    
    

\section{Methodology} 

\subsection{Preliminaries}
Spectral-domain learning~\cite{xu2020learning} has gained increasing popularity in medical image analysis~\cite{wang2022ffcnet} for its ability to identify subtle frequency patterns that may not be well detected by the pure spatial-domain network like UNet~\cite{ronneberger2015unet}.
For instance, a recent dual-encoder network, YNet~\cite{farshad2022net}, incorporates a spectral encoder with Fast Fourier Convolution (FFC)~\cite{chi2020fast} to disentangle global patterns across varying frequency components and derives hybrid feature representation. 
In addition, spectrum learning also exhibits advantageous robustness and generalization against adversarial attacks, data corruption, and distribution shifts~\cite{rao2021global}. 
In label-efficient learning, some preliminary works have been proposed to encourage mutual consistency between outputs 
from two networks~\cite{chen2021semi}, two decoders~\cite{wu2021semi}, and teacher-student models~\cite{liu2022weakly}, 
yet only in the spatial domain. As far as we know, spatial-spectral cross-domain consistency has never been investigated to promote learning with sparse annotations of medical data. 
This has motivated us to develop the cross-domain cooperative mutual teaching scheme to leverage the favorable properties when learning in the spectral space. 

Besides consistency constraints, utilizing pseudo labels as supplementary supervision is another principle in label-efficient learning~\cite{lee2020scribble2label,wang2022freematch}. 
To prevent the model from being influenced by noise and inaccuracies within the pseudo labels, numerous studies have endeavored to enhance their quality, including averaging the model predictions from several iterations~\cite{lee2020scribble2label}, filtering out unreliable pixels~\cite{wang2022freematch}, and mixing dual-branch outputs~\cite{luo2022scribble} following 
\begin{equation}
\label{eq:dmpls}
    p_{mix} = \alpha \times p_1 + (1-\alpha) \times p_2, \alpha = random(0,1),
\end{equation}
where $\alpha$ is the random mixing ratio. $p_1$, $p_2$, and $p_{mix}$ denote the probability maps from the two spatial decoders and their mixture.
These approaches only operate in the spatial domain, regardless of single or dual branches, while we consider both spatial and spectral domains and propose to adaptively merge dual-branch outputs with respective pixel-wise entropy guidance. 


\subsection{S$^2$ME: Spatial-Spectral Mutual Teaching \& Ensemble Learning}
\textbf{Spatial-Spectral Cross-domain Mutual Teaching.}
In contrast to prior weakly-supervised learning methods that have merely emphasized spatial considerations, our approach designs a dual-branch structure consisting of
a spatial branch $f_{spa}(x, \theta_{spa})$ and a spectral branch $f_{spe}(x, \theta_{spe})$, with $x$ and $\theta$ being the input image and randomly initialized model parameters.  
As illustrated in Fig.~\ref{fig:overall}, the spatial and spectral branches take the same training image as the input and extract domain-specific patterns. The raw model outputs, \textit{i.e.}, the logits $l_{spa}$ and $l_{spe}$, will be converted to probability maps $p_{spa}$ and $p_{spe}$ with \textit{Softmax} normalization, and further to respective pseudo labels $\hat{y}_{spa}$ and $\hat{y}_{spe}$ by $\hat{y} = \mathop{\arg\max}p$. 
The spatial and spectral pseudo labels supervise the other branch collaboratively during mutual teaching and can be expressed as 
\begin{equation}
\label{eq:mt}
    \hat{y}_{spa} \rightarrow f_{spe} \text{ and } \hat{y}_{spe} \rightarrow f_{spa},
\end{equation}
where ``$\rightarrow$'' denotes supervision\footnote{For convenience, we omit the input $x$ and model parameters $\theta$.}.
Through cross-domain engagement, these two branches complement each other, 
with each providing valuable domain-specific insights and feedback to the other. Consequently, such a scheme can lead to better feature extraction, more meaningful data representation, and domain-specific knowledge transmission, thus boosting model generalization and robustness. 

\noindent\textbf{Entropy-guided Pseudo Label Ensemble Learning.}
In addition to mutual teaching, we consider aggregating the pseudo labels from the spatial and spectral branches in ensemble learning, aiming to take advantage of the distinctive yet complementary properties of the cross-domain features. 
As we know, a pixel characterized by a higher entropy value indicates elevated uncertainty in terms of its corresponding prediction. 
We can observe from the entropy maps $\mathcal{H}_{spa}$ and $\mathcal{H}_{spe}$ in Fig.~\ref{fig:overall} that the pixels of the polyp boundary exhibit greater difficulties in accurate segmentation, presenting with higher entropy values (the white contours). Considering such property, 
we propose a novel adaptive strategy to automatically adjust the mixing ratio for each pixel based on the entropy of its categorical probability distribution. Hence, the mixed pseudo labels are more reliable and beneficial for ensemble learning.
Concretely, with the spatial and spectral probability maps $p_{spa}$ and $p_{spe}$, the corresponding entropy maps $\mathcal{H}_{spa}$ and $\mathcal{H}_{spe}$ can be computed with 
\begin{equation}
   \mathcal{H} = -\sum_{c=0}^{C-1} p(c) \times \log{p(c)},
\end{equation}
where $C$ is the number of classes that equals 2 in our task.
Unlike previous image-level fixed-ratio mixing or random mixing as Eq.~(\ref{eq:dmpls}), we can update the mixing ratio between the two probability maps $p_{spa}$ and $p_{spe}$ with the weighted entropy guidance at each pixel location by 
\begin{equation}
    p_{s^2} = \frac{\mathcal{H}_{spe}}{\mathcal{H}_{spa}+\mathcal{H}_{spe}} \otimes p_{spa} + \frac{\mathcal{H}_{spa}}{\mathcal{H}_{spa}+\mathcal{H}_{spe}} \otimes p_{spe},
\end{equation}
where ``$\otimes$'' denotes pixel-wise multiplication. $p_{s^2}$ is the merged probability map and can be further converted to the pseudo label by $\hat{y}_{s^2} = \mathop{\arg\max}p_{s^2}$ to supervise the spatial and spectral branch in the context of ensemble learning following
\begin{equation}
\label{eq:el}
    \hat{y}_{s^2} \rightarrow f_{spa} \text{ and } \hat{y}_{s^2} \rightarrow f_{spe}.
\end{equation}
By absorbing strengths from the spatial and spectral branches, ensemble learning from the mixed pseudo labels facilitates model optimization with reduced overfitting, increased stability, and improved generalization and robustness. 

\noindent\textbf{Hybrid Loss Supervision from Scribbles and Pseudo Labels.}
Besides the scribble annotations for partial pixels, the aforementioned three types of pseudo labels $\hat{y}_{spa}$, $\hat{y}_{spe}$, and $\hat{y}_{s^2}$ can offer complementary supervision for every pixel, with different learning regimes. Overall, our hybrid loss supervision is based on 
Cross Entropy loss $\ell_{CE}$ and Dice loss $\ell_{Dice}$. Specifically, we employ the partial Cross Entropy loss~\cite{lin2016scribblesup} $\ell_{pCE}$, which only calculates the loss on the labeled pixels, for learning from scribbles following 
\begin{equation}
    \mathcal{L}_{scrib}=\ell_{pCE}(l_{spa}, y)+\ell_{pCE}(l_{spe}, y),
\end{equation}
where $y$ denotes the scribble annotations. 
Furthermore, the mutual teaching loss with supervision from domain-specific pseudo labels is
\begin{equation}
\small
    \mathcal{L}_{mt}=\underbrace{\{\ell_{CE}(l_{spa}, \hat{y}_{spe})\!+\!\ell_{Dice}(p_{spa}, \hat{y}_{spe})\}}_{\hat{y}_{spe} \rightarrow f_{spa}} + \underbrace{\{\ell_{CE}(l_{spe}, \hat{y}_{spa})\!+\!\ell_{Dice}(p_{spe}, \hat{y}_{spa})\}}_{\hat{y}_{spa} \rightarrow f_{spe}}.
\end{equation}
Likewise, the ensemble learning loss with supervision from the enhanced mixed pseudo labels can be formulated as 
\begin{equation}
\small
    \mathcal{L}_{el}=\underbrace{\{\ell_{CE}(l_{spa}, \hat{y}_{s^2})+\ell_{Dice}(p_{spa}, \hat{y}_{s^2})\}}_{\hat{y}_{s^2} \rightarrow f_{spa}} + \underbrace{\{\ell_{CE}(l_{spe}, \hat{y}_{s^2})+\ell_{Dice}(p_{spe}, \hat{y}_{s^2})\}}_{\hat{y}_{s^2} \rightarrow f_{spe}}.
\end{equation}
Holistically, our hybrid loss supervision can be stated as
\begin{equation}\label{eq:loss_hybrid}
    \mathcal{L}_{hybrid} = \mathcal{L}_{scrib} + \lambda_{mt}\times\mathcal{L}_{mt} + \lambda_{el}\times\mathcal{L}_{el},
\end{equation}
where $\lambda_{mt}$ and $\lambda_{el}$ serve as weighting coefficients that regulate the relative significance of various modes of supervision. The hybrid loss considers all possible supervision signals in the spatial-spectral dual-branch network and exceeds partial combinations of its constituent elements, as evidenced in the ablation study.

\section{Experiments}
\subsection{Experimental Setup}

\textbf{Datasets.}
We employ the SUN-SEG~\cite{ji2022video} dataset with scribble annotations for training and assessing the in-distribution performance. This dataset is based on the SUN database~\cite{misawa2021development}, which contains 100 different polyp video cases. To reduce data redundancy and memory consumption, we choose the first of every five consecutive frames in each case. We then randomly split the data into 70, 10, and 20 cases for training, validation, and testing, leaving 6677, 1240, and 1993 frames in the respective split. 
For out-of-distribution evaluation, we utilize three public datasets, namely Kvasir-SEG~\cite{jha2020kvasir}, CVC-ClinicDB~\cite{bernal2015wm}, and PolypGen~\cite{ali2023multi} with 1000, 612, and 1537 polyp frames, respectively. These datasets are collected from diversified patients in multiple medical centers with various data acquisition systems. Varying data shifts and corruption like motion blur and specular reflections\footnote{Some exemplary polyp frames are presented in the supplementary materials.} pose significant challenges to model generalization and robustness.


\noindent\textbf{Implementation Details.}
We implement our method with PyTorch~\cite{paszke2017automatic} and run the experiments on a single NVIDIA RTX3090 GPU. 
The SGD optimizer is utilized for training 30k iterations with a momentum of 0.9, a weight decay of 0.0001, and a batch size of 16. The execution time for each experiment is approximately 4 hours.
The initial learning rate is 0.03 and updated with the poly-scheduling policy~\cite{luo2022scribble}. 
The loss weighting coefficients $\lambda_{mt}$ and $\lambda_{el}$ are empirically set the same and exponentially ramped up~\cite{chen2021semi} from 0 to 5 in 25k iterations.
All the images are randomly cropped at the border with maximally 7 pixels and resized to $224 \times 224$ in width and height. Besides, random horizontal and vertical flipping are applied with a probability of 0.5, respectively. 

We utilize UNet~\cite{ronneberger2015unet} and YNet~\cite{farshad2022net} as the respective segmentation model in the spatial and spectral branches.
The performance of the scribble-supervised model with partial Cross Entropy~\cite{lin2016scribblesup} loss (Scrib-pCE) and the fully-supervised model with Cross Entropy loss (Fully-CE) are treated as the lower and upper bound, respectively. 
Five classical and relevant methods, including EntMin~\cite{grandvalet2004semi}, GCRF~\cite{obukhov2019gated}, USTM~\cite{liu2022weakly}, CPS~\cite{chen2021semi}, and DMPLS~\cite{luo2022scribble} are employed as the comparative baselines and implemented with UNet~\cite{ronneberger2015unet} as the segmentation backbone referring to the WSL4MIS\footnote{\url{https://github.com/HiLab-git/WSL4MIS}} repository. 
For a fair comparison, the output from the spatial branch is taken as the final prediction and utilized in evaluation without post-processing. In addition, statistical evaluations are conducted with multiple seeds, and the mean and standard deviations of the results are reported. 



\subsection{Results and Analysis}

\begin{table}[!h]
  \centering
  \caption{Quantitative comparison of the in-distribution segmentation performance. 
  The shaded grey and blue rows are the lower and upper bound. The best results of the scribble-supervised methods are in bold. }
  \resizebox{0.73\textwidth}{!}{%
    \begin{tabular}{ccccc}
    \hline
    \multirow{2}[2]{*}{Method} & \multicolumn{4}{c}{SUN-SEG~\cite{ji2022video}} \\
    \cline{2-5}
          & DSC $\uparrow$ & IoU $\uparrow$   & Prec $\uparrow$ & HD $\downarrow$ \\
    \hline
   \rowcolor[HTML]{DCDCDC} Scrib-pCE~\cite{lin2016scribblesup} & 0.633±0.010 & 0.511±0.012 & 0.636±0.021 & 5.587±0.149 \\
    EntMin~\cite{grandvalet2004semi} & 0.642±0.012 & 0.519±0.013 & 0.666±0.016 & 5.277±0.063 \\
    GCRF~\cite{obukhov2019gated}  & 0.656±0.019 & 0.541±0.022 & 0.690±0.017 & 4.983±0.089 \\
    USTM~\cite{liu2022weakly}  & 0.654±0.008 & 0.533±0.009 & 0.663±0.011 & 5.207±0.138 \\
    CPS~\cite{chen2021semi}   & 0.658±0.004 & 0.539±0.005 & 0.676±0.005 & 5.092±0.063 \\
    DMPLS~\cite{luo2022scribble} & 0.656±0.006 & 0.539±0.005 & 0.659±0.011 & 5.208±0.061 \\
    \textbf{S$^2$ME (Ours)}  & \textbf{0.674±0.003} & \textbf{0.565±0.001} & \textbf{0.719±0.003} & \textbf{4.583±0.014} \\
    \rowcolor[HTML]{DAE8FC} Fully-CE & 0.713±0.021 & 0.617±0.023 & 0.746±0.027 & 4.405±0.119 \\
    \hline
    \end{tabular}%
    }
  \label{tab:overall_quant}%
\end{table}%

\begin{figure}[!ht]
\centering
\includegraphics[width=0.9\textwidth]{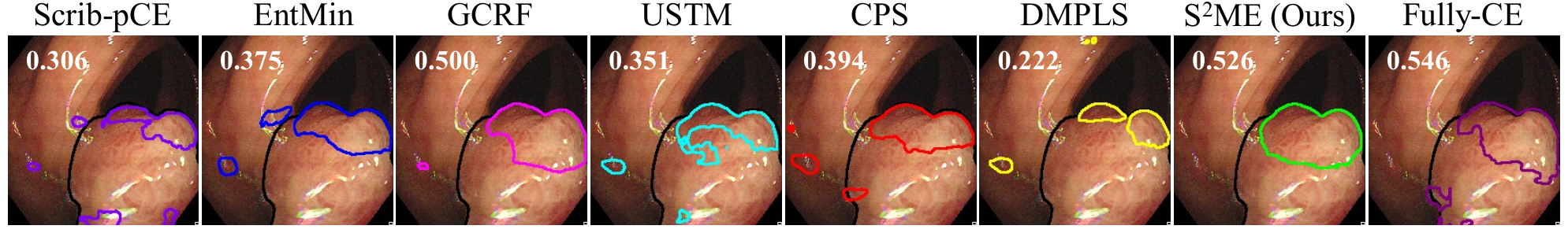}
\caption{Qualitative performance comparison one camouflaged polyp image with DSC values on the left top. The contour of the ground-truth mask is displayed in black, in comparison with that of each method shown in different colors.}
\label{fig:quality}
\centering
\end{figure}

The performance of weakly-supervised methods is assessed with four metrics,\textit{i.e.}, Dice Similarity Coefficient (DSC), Intersection over Union (IoU), Precision (Prec), and a distance-based measure of Hausdorff Distance (HD). 
As shown in Table~\ref{tab:overall_quant} and Fig.~\ref{fig:quality}, our S$^2$ME achieves superior in-distribution performance quantitatively and qualitatively compared with other baselines on the SUN-SEG~\cite{ji2022video} dataset. 
Regarding generalization and robustness, as indicated in Table~\ref{tab:gen}, our method outperforms other weakly-supervised methods by a significant margin on three unseen datasets, 
and even exceeds the fully-supervised upper bound on two of them\footnote{Complete results of all four metrics are present in the supplementary materials.}. These results suggest the efficacy and reliability of the proposed solution S$^2$ME 
in fulfilling polyp segmentation tasks with only scribble annotations. Notably, the encouraging performance on unseen datasets exhibits promising clinical implications in deploying our method to real-world scenarios.

\begin{table}[!ht]
  \centering
  \caption{Generalization comparison on three unseen datasets. The underlined results surpass the upper bound.}
  \resizebox{0.92\textwidth}{!}{%
    \begin{tabular}{ccccccc}
    \hline
    \multirow{2}[2]{*}{Method} & \multicolumn{2}{c}{Kvasir-SEG~\cite{jha2020kvasir}} & \multicolumn{2}{c}{CVC-ClinicDB~\cite{bernal2015wm}} & \multicolumn{2}{c}{PolypGen~\cite{ali2023multi}} \\
    \cline{2-7}
          & DSC $\uparrow$  & HD $\downarrow$   & DSC $\uparrow$  & HD $\downarrow$    & DSC $\uparrow$  & HD $\downarrow$ \\
    \hline
    \rowcolor[HTML]{DCDCDC} Scrib-pCE~\cite{lin2016scribblesup} & 0.679±0.010 & 6.565±0.173 & 0.573±0.016 & 6.497±0.156 & 0.524±0.012 & 6.084±0.189 \\
    EntMin~\cite{grandvalet2004semi} & 0.684±0.004 & 6.383±0.110 & 0.578±0.016 & 6.308±0.254 & 0.542±0.003 & 5.887±0.063 \\
    GCRF~\cite{obukhov2019gated}  & 0.702±0.004 & 6.024±0.014 & 0.558±0.008 & 6.192±0.290 & 0.530±0.006 & 5.714±0.133 \\
    USTM~\cite{liu2022weakly}  & 0.693±0.005 & 6.398±0.138 & 0.587±0.019 & 5.950±0.107 & 0.538±0.007 & 5.874±0.068 \\
    CPS~\cite{chen2021semi}   & 0.703±0.011 & 6.323±0.062 & 0.591±0.017 & 6.161±0.074 & 0.546±0.013 & 5.844±0.065 \\
    DMPLS~\cite{luo2022scribble} & 0.707±0.006 & 6.297±0.077 & 0.593±0.013 & 6.194±0.028 & 0.547±0.007 & 5.897±0.045 \\
    \textbf{S$^2$ME (Ours)} & \textbf{0.750±0.003} & \textbf{5.449±0.150} & \underline{\textbf{0.632±0.010}} & \underline{\textbf{5.633±0.008}} & \underline{\textbf{0.571±0.002}} & \underline{\textbf{5.247±0.107}} \\
    \rowcolor[HTML]{DAE8FC} Fully-CE & 0.758±0.013 & 5.414±0.097 & 0.631±0.026 & 6.017±0.349 & 0.569±0.016 & 5.252±0.128 \\
    \hline
    \end{tabular}%
    }
  \label{tab:gen}%
\end{table}%

\subsection{Ablation Studies}
\textbf{Network Structures.}
We first conduct the ablation analysis on the network components. As shown in Table~\ref{tab:abl_net}, the spatial-spectral configuration of our S$^2$ME yields superior performance compared to single-domain counterparts with ME, confirming the significance of utilizing cross-domain features. 
\begin{table}[!h]
  \centering
  \caption{Ablation comparison of dual-branch network architectures. Results are from outputs of Model-1 on the SUN-SEG~\cite{ji2022video} dataset.}
    \resizebox{0.88\textwidth}{!}{%
    \begin{tabular}{ccccccc}
    \hline
    Model-1 & Model-2 & Method & DSC $\uparrow$  & IoU $\uparrow$   & Prec $\uparrow$  & HD $\downarrow$ \\
    \hline
    UNet~\cite{ronneberger2015unet}  & UNet~\cite{ronneberger2015unet}  & \multicolumn{1}{c}{\multirow{2}[0]{*}{ME (Ours)}} & 0.666±0.002 & 0.557±0.002 & 0.715±0.008 & 4.684±0.034 \\
    YNet~\cite{farshad2022net}  & YNet~\cite{farshad2022net}  &       & 0.648±0.004 & 0.538±0.005 & 0.695±0.004 & 4.743±0.006 \\
    UNet~\cite{ronneberger2015unet}  & YNet~\cite{farshad2022net}  & S$^2$ME (Ours) & \textbf{0.674±0.003} & \textbf{0.565±0.001} & \textbf{0.719±0.003} & \textbf{4.583±0.014} \\
    \hline
    \end{tabular}%
    }
  \label{tab:abl_net}%
\end{table}%

\noindent\textbf{Pseudo Label Fusion Strategies.} 
To ensure the reliability of the mixed pseudo labels for ensemble learning, we present the pixel-level adaptive fusion strategy according to entropy maps of dual predictions to balance the strengths and weaknesses of spatial and spectral branches. 
As demonstrated in Table~\ref{tab:abl_fusion}, our method achieves improved performance compared to two image-level fusion strategies, \textit{i.e.}, random~\cite{luo2022scribble} and equal mixing. 

\begin{table}[!h] \centering
\begin{minipage}[!h]{0.48\linewidth}
    \caption{Ablation on the pseudo label fusion strategies on the SUN-SEG~\cite{ji2022video} dataset.} \label{tab:abl_fusion}
    \resizebox{\textwidth}{!}{%
    \begin{tabular}{cccccc}
    \hline
    \multicolumn{2}{c}{Fusion} & \multicolumn{2}{c}{Metrics} \\
    \hline
    Strategy & Level & DSC $\uparrow$  & HD $\downarrow$\\
    \hline
    Random~\cite{luo2022scribble} & Image & 0.665±0.008  & 4.750±0.169 \\
    Equal (0.5) & Image & 0.667±0.001 & 4.602±0.013 \\
    Entropy (Ours) & Pixel & \textbf{0.674±0.003}  & \textbf{4.583±0.014} \\
    \hline
    \end{tabular}%
    }
\end{minipage}\hfill
    \begin{minipage}[!h]{0.48\linewidth}
    \caption{Ablation study on the loss components on the SUN-SEG~\cite{ji2022video} dataset.} \label{tab:abl_loss}
    \resizebox{\textwidth}{!}{
    \begin{tabular}{p{0.9cm}<{\centering}p{0.9cm}<{\centering}p{0.9cm}<{\centering}cccc}
    \hline
    \multicolumn{3}{c}{Loss} & \multicolumn{2}{c}{Metrics} \\
    \hline
    $\mathcal{L}_{scrib}$ & $\mathcal{L}_{mt}$  & $\mathcal{L}_{el}$ & DSC $\uparrow$ & HD  $\downarrow$\\
    \hline
    \ding{51} & \ding{55} & \ding{55} & 0.627±0.004 & 5.580±0.112 \\
    \ding{51} & \ding{51} & \ding{55} & 0.668±0.007 & 4.782±0.020 \\
    \ding{51} & \ding{55} & \ding{51} & 0.662±0.004 & 4.797±0.146 \\
    \ding{51} & \ding{51} & \ding{51} & \textbf{0.674±0.003} & \textbf{4.583±0.014} \\
    \hline
    \end{tabular}%
    }
\end{minipage}
\end{table}


\noindent\textbf{Hybrid Loss Supervision.}
We decompose the proposed hybrid loss $\mathcal{L}_{hybrid}$ in Eq.~(\ref{eq:loss_hybrid}) to demonstrate the effectiveness of holistic supervision from scribbles, mutual teaching, and ensemble learning. As shown in Table~\ref{tab:abl_loss}, our proposed hybrid loss, involving $\mathcal{L}_{scrib}$, $\mathcal{L}_{mt}$, and $\mathcal{L}_{el}$, achieves the optimal results.


\section{Conclusion}
To our best knowledge, we propose the first spatial-spectral dual-branch network structure for weakly-supervised medical image segmentation that efficiently leverages cross-domain patterns with collaborative mutual teaching and ensemble learning. 
Our pixel-level entropy-guided fusion strategy advances the reliability of the aggregated pseudo labels, which provides valuable supplementary supervision signals.
Moreover, we optimize the segmentation model with the hybrid mode of loss supervision from scribbles and pseudo labels in a holistic manner and witness improved outcomes. 
With extensive in-domain and out-of-domain evaluation on four public datasets, our method shows superior accuracy, generalization, and robustness, indicating its clinical significance in alleviating data-related issues such as data shift and corruption which are commonly encountered in the medical field. 
Future efforts can be paid to apply our approach to other annotation-efficient learning contexts like semi-supervised learning, other sparse annotations like points, and more medical applications. 

\subsubsection{Acknowledgements.}

This work was supported by the Shun Hing Institute of Advanced Engineering (SHIAE project BME-p1-21) at the Chinese University of Hong Kong
(CUHK), Hong Kong Research Grants Council (RGC) Collaborative Research
Fund (CRF C4026-21GF and CRF C4063-18G), General Research Fund (GRF
14216022), Shenzhen-Hong Kong-Macau Technology Research Programme
(Type C) STIC Grant SGDX20210823103535014 (202108233000303), and
(GRS) \#3110167.

\bibliographystyle{splncs04}
\bibliography{references}

\end{document}